\documentclass[prb,reprint,superscriptaddress]{revtex4-2}
 
\usepackage{hyperref}
\usepackage{amsmath,amssymb,amsfonts}
\usepackage{graphicx}
\usepackage{textcomp}
\usepackage{algorithm}
\usepackage{multirow}
\usepackage{algpseudocode}
\usepackage{xcolor}
\usepackage{makecell}
\usepackage{kbordermatrix}

\begin{document}

\title{TensorMD: Scalable Tensor-Diagram based Machine Learning Interatomic Potential on Heterogeneous Many-Core Processors}


\author{Xin Chen}
\thanks{These two authors contributed equally}
\affiliation{Institute of Applied Physics and Computational Mathematics, Beijing, China}
\author{Yucheng Ouyang}
\thanks{These two authors contributed equally}
\affiliation{SKL of Computer Architecture, Institute of Computing Technology, Chinese Academy of Sciences, Beijing, China}
\author{Xin Chen}
\affiliation{National Research Center of Parallel Computer Engineering and Technology, Beijing, China}

 
\author{Zhenchuan Chen}
\affiliation{SKL of Computer Architecture, Institute of Computing Technology, Chinese Academy of Sciences, Beijing, China}

\author{Rongfen Lin}
\affiliation{National Research Center of Parallel Computer Engineering and Technology, Beijing, China}

\author{Xingyu Gao}
\affiliation{Institute of Applied Physics and Computational Mathematics, Beijing, China}

\author{Lifang Wang}
\affiliation{Institute of Applied Physics and Computational Mathematics, Beijing, China}

\author{Fang Li}
\affiliation{National Research Center of Parallel Computer Engineering and Technology, Beijing, China}

\author{Yin Liu}
\affiliation{SKL of Computer Architecture, Institute of Computing Technology, Chinese Academy of Sciences, Beijing, China}

\author{Honghui Shang}
\email[Corresponding author:]{shanghonghui@ict.ac.cn}
\affiliation{SKL of Computer Architecture, Institute of Computing Technology, Chinese Academy of Sciences, Beijing, China}

\author{Haifeng Song}
\email[Corresponding author:]{song_haifeng@iapcm.ac.cn}
\affiliation{Institute of Applied Physics and Computational Mathematics, Beijing, China}


%
%

\begin{abstract}
Molecular dynamics simulations have emerged as a potent tool for investigating 
the physical properties and kinetic behaviors of materials at the atomic scale,
particularly in extreme conditions. Ab initio accuracy is now achievable with 
machine learning based interatomic potentials. With recent advancements in 
high-performance computing, highly accurate and large-scale simulations become 
feasible. This study introduces TensorMD, a new machine learning interatomic 
potential (MLIP) model that integrates physical principles and tensor diagrams. 
The tensor formalism provides a more efficient computation and greater 
flexibility for use with other scientific codes. Additionally, we proposed 
several portable optimization strategies and developed a highly optimized 
version for the new Sunway supercomputer. Our optimized TensorMD can achieve 
unprecedented performance on the new Sunway, enabling simulations of up to 52 
billion atoms with a time-to-solution of 31 ps/step/atom, setting new records 
for HPC + AI + MD.
\end{abstract}

\maketitle

%
%

\section{Introduction}

Atomistic simulation is a powerful theoretical method used to study the physical 
and chemical properties of materials at the atomic scale. Among these methods, 
molecular dynamics simulation has become widely used in studying the physical 
properties of many metal materials and modeling multiphase equations of states. 
However, the accuracy of molecular dynamics simulation depends on the atomic
potential that describes the interaction between atoms. 

First-principles methods are highly accurate but computationally expensive. 
On the other hand, semi-empirical potentials are computationally efficient, 
but their accuracy is usually not sufficient for quantitative research. 
In recent years, the development of atomic potentials based on machine 
learning methods has gained widespread attention. These machine learning 
interatomic potentials (MLIPs) can approach the accuracy of first-principles 
methods while remaining computationally efficient, making them suitable for 
large-scale simulations, even at experimentally observable scales. 

MLIPs have been used to study the dynamic behavior and extreme physical 
properties of materials realistically and accurately. For example, Oganov 
constructed the phase diagram of Uranium\cite{Oganov_U} in a wide 
temperature-pressure range with MLIPs. Similarly, Zong et al\cite{Zong_Zr}. 
studied the martensitic phase transition of Zr based on a modified 
Behler-Parinello model, while William et al. studied the extreme physical 
properties of high-density carbon using SNAP 
potential\cite{snap_gb_sc21,snap_hdc_prb}, which directly served the inertial 
confinement fusion research. By using these powerful techniques, scientists 
can now investigate the behavior of materials at the atomic scale with a high 
level of accuracy and efficiency, opening up new avenues for research and 
development. Recently, many traditional first-principles programs, such as 
VASP\cite{vasp_mlff_1,vasp_mlff_2}, have introduced MLIPs based acceleration 
modules, which can significantly improve the computational efficiency of 
\textit{ab initio} molecular dynamics simulations.

Until now, various MLIPs have been proposed, including GAP\cite{soap0,soap1,soap2,soap3}, SNAP\cite{SNAP,SNAP_Algo,snap_gb_sc21,SNAP_New}, MLFF\cite{vasp_mlff_1,vasp_mlff_2}, HDNNP (BP)\cite{BPNN}, DP\cite{DP_GB,DPMD_PRL,DP1,DP_Ppopp,DP_compress} et al. These models improve the accuracy and computational efficiency of atomic potentials to various degrees, making them important tools in materials computational simulation. Moreover, SNAP and DP have developed high-performance versions that leverage supercomputers such as Summit\cite{snap_gb_sc21,DP_GB,DP_Ppopp} and Fugaku\cite{DP_Ppopp}. These versions enable molecular dynamics simulations of billions of atoms and provide powerful tools for studying the large-scale dynamic behavior and extreme physical properties of materials.

However, most previous high-performance MLIP works have focused on GPUs, with relatively less attention given to many-core processors. While DP is the first optimized potential for many-core processors\cite{DP_Ppopp}, it is worth noting that Fugaku, which is built with exceptionally fast memory, may not be representative of other HPC systems. Another highly regarded supercomputer family is Sunway, which includes Sunway TaihuLight and its successor, the new generation of Sunway. Sunway TaihuLight held the top spot on the TOP500 list from 2016 to 2018, and it was known for its high computational power and energy efficiency. The new generation of Sunway is also expected to be a powerful machine and is likely to be used for a wide range of scientific applications. While there has been a lot of previous works to improve molecular dynamics simulations on Sunway TaihuLight\cite{MD_SW_Taihulight_1,MD_SW_Taihulight_2} and the new generation of Sunway\cite{LMFF}, these efforts have mainly focused on empirical potentials, and there has been relatively little research on MLIP.

Therefore, in this work, we proposed a new MLIP with a comprehensive tensor formalism, TensorMD, and optimized it for the new generation Sunway supercomputer. Our approach demonstrated exceptional performance, allowing for the simulation of 52 billion atoms with a time-to-solution of 3.1 ps/step/atom on the full machine. This sets new records for HPC + AI molecular dynamics, and highlights the potential of TensorMD for accelerating scientific research in materials science.

The work presented in this paper makes several significant contributions to the field of machine learning interactomic potential:

\begin{itemize}
    \item[1.] We introduce a comprehensive and straightforward tensor formalism for representing these potentials. Our approach simplifies the representation of interactomic potentials and enables greater flexibility for model development.
    \item[2.] Our implementation is based on BLAS and achieves very high efficiency. The lack of dependency on deep learning backends allows for easier integration with other scientific codes.
    \item[3.] We propose a series of portable optimization techniques that significantly improve the efficiency of our implementation on the new Sunway supercomputer.
    \item[4.] Finally, our scaling tests demonstrate that TensorMD can simulate up to 52 billion atoms with a time-to-solution of 3.1 ps/step/atom, setting new records for HPC + AI + MD. These results show that our approach can be used to solve large-scale molecular dynamics problems on the new Sunway supercomputer, which is an important step forward in this field.
\end{itemize}

%
%

\section{Background}
\subsection{The Basic Formalism of TensorMD}
\label{sec:nnp}

The construction of a neural network interaction potential typically involves two models: the atomic feature model, which characterizes atomic environments, and the regression model, which maps atomic features to atomic energies. The total energy of a system is the sum of atomic energies. The atomic feature model must satisfy three fundamental invariants, namely translation, rotation, and permutation. A widely used method to construct atomic features is through the utilization of atomic descriptor functions. These functions can be divided into two categories\cite{BPNN}: radial functions, which explain atom-atom interactions, and angular functions, which describe triple-atom interactions. The general forms of these functions are as follows:
\begin{eqnarray}
  \label{eqn:general_atomic_descriptor_functions}
G_i^{\mathrm{radial}} & = & \sum_{j}^{N}{f(r_{ij})} \\
G_i^{\mathrm{angular}} & = & \sum_{j}^{N}{\sum_{k \ne j}^{N}}{P_{m}(\cos{\theta_{jik}})f(r_{ij})f(r_{ik})}
\end{eqnarray}
where $i$ is the center atom, $\theta_{jik}$ is the angle formed by atoms $j$, $i$ and $k$, and $P_{m}(\cos{\theta_{jik}})$ is the $m$-th order polynomial of $\theta_{jik}$. Radial interactions can be computed with ease and efficiency, whereas the computational cost of the angular interactions increases quadratically with the number of neighbors due to its double summation requirement. Nevertheless, Daw and Baskes demonstrated that this double summation can be simplified to a single summation while keeping the three basic invariants\cite{Baskes_MEAM_1989,Daw_1989}:
\begin{eqnarray}
\label{eqn:double_to_single_simplification}
    \cos{\theta_{jik}} = \frac{r_{ij} \cdot r_{ik}}{|r_{ij}||r_{ik}|} = \frac{\sum_{\alpha\beta}{r_{ij}^{\alpha}r_{ik}^{\beta}}}{|r_{ij}||r_{ik}|}
\end{eqnarray}
where $\alpha$, $\beta$ are Cartesian directions. By substituting \ref{eqn:double_to_single_simplification} into \ref{eqn:general_atomic_descriptor_functions}, Baskes derived his MEAM formalism\cite{Baskes_MEAM_1992}. The partial electron density functions are as follows:
\begin{eqnarray}
\label{eqn:moment_s_formalism}
    \sum_{j \ne k}f(r_{ij})f(r_{ik})\cos^{0}{\theta} & = &\left[\sum_{j}f(r_{ij})\right]^2 \\
\sum_{j \ne k}f(r_{ij})f(r_{ik})\cos^{1}\theta
& = & \sum_{\alpha}\left[\sum_{j}\frac{r_{ij}^\alpha}{r_{ij}}f(r_{ij})\right]^2 \\
\sum_{j \ne k}f(r_{ij})f(r_{ik})\cos^{2}\theta
& = & \sum_{\alpha,\beta}\left[\sum_{j}
\frac{r_{ij}^\alpha r_{ij}^\beta}{r_{ij}^2}f(r_{ij})\right]^2 \\
\sum_{j \ne k}f(r_{ij})f(r_{ik})\cos^{3}\theta
& = &\sum_{\alpha,\beta,\gamma}\left[\sum_{j}
\frac{r_{ij}^\alpha r_{ij}^\beta r_{ij}^\gamma}{r_{ij}^3}f(r_{ij})\right]^2
\end{eqnarray}

Baskes suggests that the properties of atoms can be understood by considering the contributions of different types of angular momentum (spdf) to the overall electron density around the atom\cite{Pu_MEAM_2000}. Specifically, the electron density associated with each type of angular momentum is related to an average of the actual atomic electron densities that depend on that type of angular momentum. In other words, the angular momentum-dependent electron densities of individual electrons in an atom are averaged to give a general picture of the electron density distribution around the atom. Ionov and Dremov further extends the MEAM formalism and proposed a generalized embedded atom method (GEAM)\cite{GEAM}. 

Inspired by these works, the basic atomic feature descriptor functions of our TensorMD are as follows:
\begin{eqnarray}
\label{eqn:original_tensormd_formalism}
    g^{(m,k)}_{i}=\sum_{\alpha,\beta,\gamma\dots}n^{\alpha\beta\gamma\dots}
\left(\sum_{j}^{N}\frac{r_{ij}^\alpha r_{ij}^\beta r_{ij}^\gamma \dots}{r_{ij}^m} f_{k}(r_{ij})f_c(r_{ij})\right)^2
\end{eqnarray}
where $f_c(r_{ij})$ is a cutoff function. $m$ can be considered as the angular moment. Equation \ref{eqn:moment_s_formalism} suggests that the radial interaction is the square of $g^{(0, k)}$.


\subsection{HPC Platforms and Software Environment}
In this study, performance assessment was conducted using the new-generation Sunway supercomputer, which is the successor to the Sunway TaihuLight supercomputer. Similar to its predecessor, the new Sunway system utilizes high-performance heterogeneous many-core processors and interconnection network chips that are domestically developed in China. The SW39000 processor of the new Sunway supercomputer is specifically designed to handle massive thread and data parallelism, resulting in exceptional performance on parallel workloads. Each processor comprises six core groups (CGs) and each CG contains 65 cores, resulting in a total of 390 cores. Each CG consists of a management processing element (MPE), a cluster of computing processing elements (CPEs), a memory controller (MC) and 16 GB DDR4 memory (12 channels) with theoretical bandwidth 51.2 GB/s. The MPE is responsible for computations, management, and communication within each CG, while the CPEs are structured as an $8\times 8$ mesh (64 cores) to maximize aggregated computing power and minimize micro-architecture complexity. The frequency of a CPE is 2.15 GHz. Each CPE has 256 KB fully-controllable local device memory (LDM). CPEs support 512-bit single instruction multiple data (SIMD) operations. The CPEs employ a mesh network to enable high-bandwidth data communication (P2P and collective communications) among CPEs within a CG, which is known as remote scratchpad memory access (RMA). 

%

\section{Innovation and Optimization}

%

\subsection{The Tensor Formalism}
\label{sec:tensor_formalism}

Although Equation \ref{eqn:original_tensormd_formalism} is a comprehensive formalism, it still involves numerous for loops, making it complicated to handle. 
Building on the success of TensorAlloy implementation of the embedded atom method and angular-dependent potential within a TensorFlow framework\cite{TensorAlloy_2}, we have developed a complete tensor formalism for our potential. 

\begin{figure}
  \begin{center}
  \centering
  \includegraphics[width=0.4\textwidth]{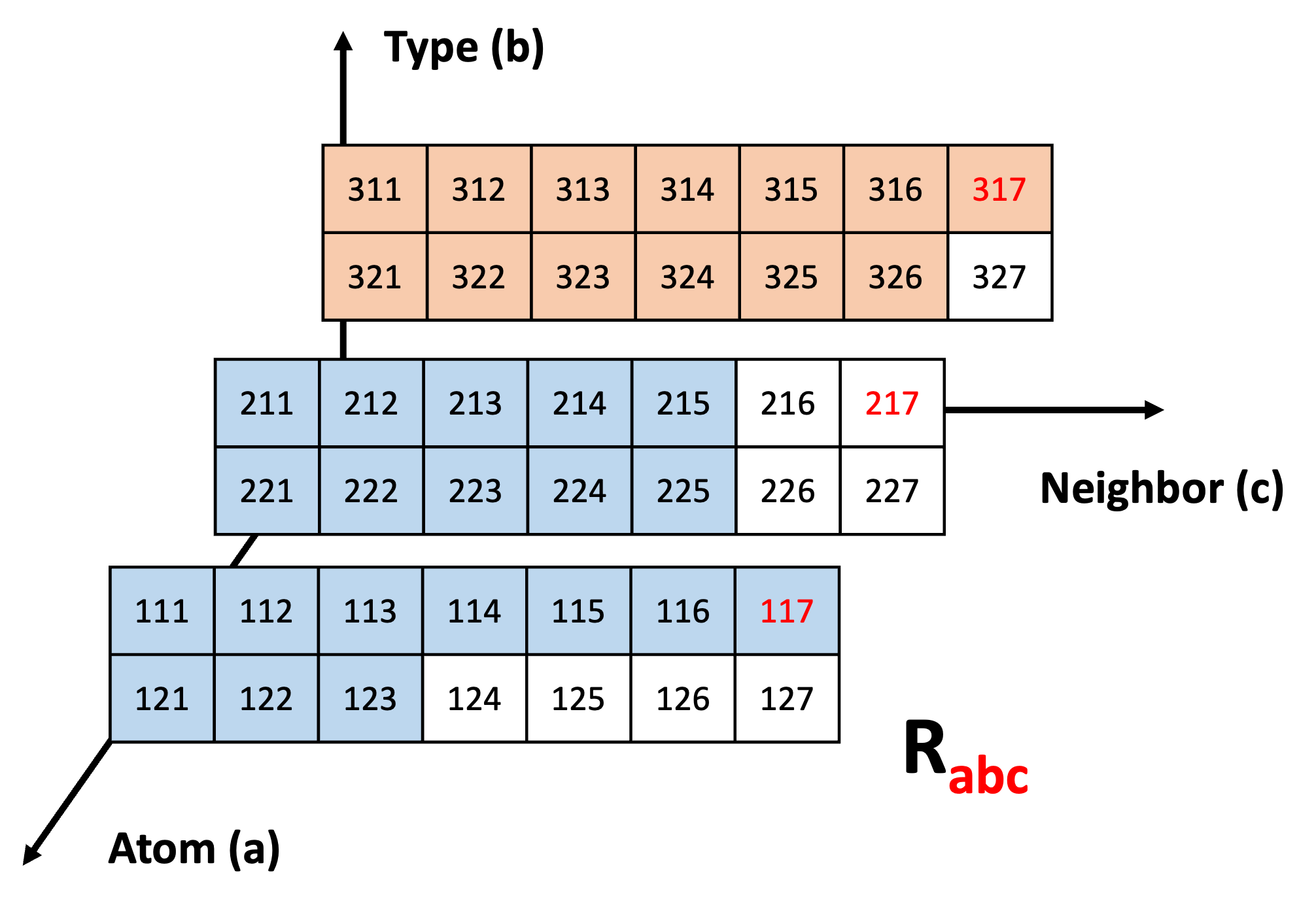}
  \caption{\label{fig:R_abc} The view of the $R_{abc}$ tensor. }
  \end{center}
\end{figure}

Figure \ref{fig:R_abc} illustrates how the neighbor table of a multi-atomic system can be represented by a three-dimensional matrix. This matrix is a third-order tensor denoted as $R_{abc}$, where $a$ represents the central atom number, $b$ represents the atom pair type number, and $c$ represents the neighbor number. For example, $R_{117}$ represents the distance between atom $1$ and its 7th type-1 neighbor. However, in actual systems, the number of neighbors for each atom may vary. If the number of neighbors of a particular type $c_i$ for an atom is less than $c_{max}$, then the values after $c_i$ can be set to zero, represented by the white grid points in Figure \ref{fig:R_abc}. To indicate whether $(a,b,c)$ is a real neighbor or zero-padded, we use a third-order tensor denoted as $mask_{abc}$.

Similarly, we can construct a fourth-order distance component tensor, denoted as $D_{abcx}$, and its reduced form, $D_{abcx}$, where $x$ represents the three directions of the Cartesian coordinate system. 

\begin{eqnarray}
    R_{abc} = \sqrt{\left( \sum_{x=0}^{x<3}{\hat{D}_{abcx}} \right)^2} \\
    \hat{D}_{abcx} = D_{abcx}R_{abc}
\end{eqnarray}

Assume $h^{(k)}(r_{ij})=f_k(r_ij)f_c(r_ij)$ and apply all $k$ different radial descriptor functions, $h^k(\cdot)$, to $R_{abc}$, we obtain the fourth-order tensor $H_{abck}$ that characterizes radial interactions. The radial fingerprint of the atom can be obtained directly through summation of the elements of $H_{abck}$.

\begin{eqnarray}
    \label{eqn:G_radial_tensor_form}
    G_{abk} = \sum_{c}{H_{abck}}
\end{eqnarray}

The matrix $G_{abk}$ in Equation \ref{eqn:G_radial_tensor_form} can be viewed as a matrix with $a$ rows and $b \times k$ columns. If the descriptor functions $f_{k}$ were selected to be Gaussian functions, then $G_{abk}$ would be equivalent to the feature matrix of the BPNN (only considering radial interactions)\cite{BPNN,TensorAlloy_1}.

To include angular interactions in our tensor formalism, we first generate the multiple moment tensor $M_{abcd}$ based on the third-order tensor $R_{abc}$ and the fourth-order distance component tensor $D_{abcx}$. The length of dimension $d$ is determined by the maximum angular momentum $m$ and can be simplified by symmetry. To represent the multiplicity of different moments, we use a second-order tensor $T_{dm}$, as shown in Equation \ref{eqn:T_dm}.

\renewcommand{\kbldelim}{(}
\renewcommand{\kbrdelim}{)}
\begin{equation}
\label{eqn:T_dm}
  T_{dm} = \kbordermatrix{
        & m = 0 & m = 1 & m = 2 & m = 3 \\
        & 1     & 0     & 0     & 0     \\
    x   & 0     & 1     & 0     & 0     \\
    y   & 0     & 1     & 0     & 0     \\
    z   & 0     & 1     & 0     & 0     \\
    xx  & 0     & 0     & 1     & 0     \\
    xy  & 0     & 0     & 2     & 0     \\
    xz  & 0     & 0     & 2     & 0     \\
    yy  & 0     & 0     & 1     & 0     \\
    yz  & 0     & 0     & 2     & 0     \\
    zz  & 0     & 0     & 1     & 0     \\
    xxx & 0     & 0     & 0     & 1     \\
    xxy & 0     & 0     & 0     & 3     \\
    xxz & 0     & 0     & 0     & 3     \\
    xyy & 0     & 0     & 0     & 3     \\
    xyz & 0     & 0     & 0     & 6     \\
    xzz & 0     & 0     & 0     & 3     \\
    yyy & 0     & 0     & 0     & 1     \\
    yyz & 0     & 0     & 0     & 3     \\
    yzz & 0     & 0     & 0     & 3     \\
    zzz & 0     & 0     & 0     & 1     \\
  }    
\end{equation}

With these tensors, we can derive the straightforward tensor representation of atomic features (Equation \ref{eqn:original_tensormd_formalism}) as:
\begin{eqnarray}
\label{eqn:tensor_formalism}
G_{abkm} = \sigma\left(T_{dm}\left(H_{abck}M_{abcd}\right)^2\right)
\end{eqnarray}
where $\sigma$ is a special function operating on a fourth-order tensor:
\begin{eqnarray}
    \label{eqn:sigma_function}
    \sigma(G[:, :, :, m]) = \begin{cases}
        \sqrt{G[:, :, :, m]} & m = 0\\
        G[:, :, :, m] & m > 0
    \end{cases}
\end{eqnarray}

In this formalism, $G_{abkm}$ is a matrix of $a$ rows and $b \times k \times (m + 1)$ columns, with each row representing the atomic features of the corresponding atoms. Passing $G_{abkm}$ into the neural network energy model (or other regression models), we can obtain the total energy $E$, the atomic energy tensor $E_{a}$ and atomic forces $F_{ax}$, as follows:

\begin{eqnarray}
E_{a} & = & \mathbf{NNP}(G_{abkm}) \\
E & = & \sum_{a}{E_{a}} \\
F_{ax} & = & -G^{\prime}_{abkm} \cdot \frac{\partial G_{abkm}}{\partial {r_{ax}}}
\end{eqnarray}

The tensor formalism also simplifies the computation of atomic forces, as we can directly obtain $G^{\prime}_{abkm}$, the derivatives of $E$ with respect to $G_{abkm}$, using standard neural network backpropagation algorithms. 

Assume $P_{abkd} = M_{abcd} H_{abck}$ and $S_{abdk} = (P_{abkd})^2$ then:
\begin{eqnarray}
    \label{eqn:dP_abkd}
    P_{abkd}^{\prime} & = & 2 T_{dm}\sigma_{abkm}^{\prime} P_{abkd} \\\
    \label{eqn:U_abck}
    U_{abck} & = & P_{abkd}^{\prime} M_{abcd} \\
    V_{abcd} & = & P_{abkd}^{\prime} H_{abck} \\
    \label{eqn:F1_abcx}
    F^{(1)}_{abcx} & = & U_{abck} H^{\prime}_{abck} R^{\prime}_{abcx} \\
    \label{eqn:F2_abcx}
    F^{(2)}_{abcx} & = & V_{abcd} M^{\prime}_{abcxd}
\end{eqnarray}
and the final atomic forces $F_{ax}$ can be obtained with a special scattered summation operation:
\begin{eqnarray}
    F_{ax} = \mathrm{ScatterSum}{\left(F^{(1)}_{abcx} + F^{(2)}_{abcx}\right)}
\end{eqnarray}

\begin{figure*}
  \begin{center}
  \centering
  \includegraphics[width=0.9\textwidth]{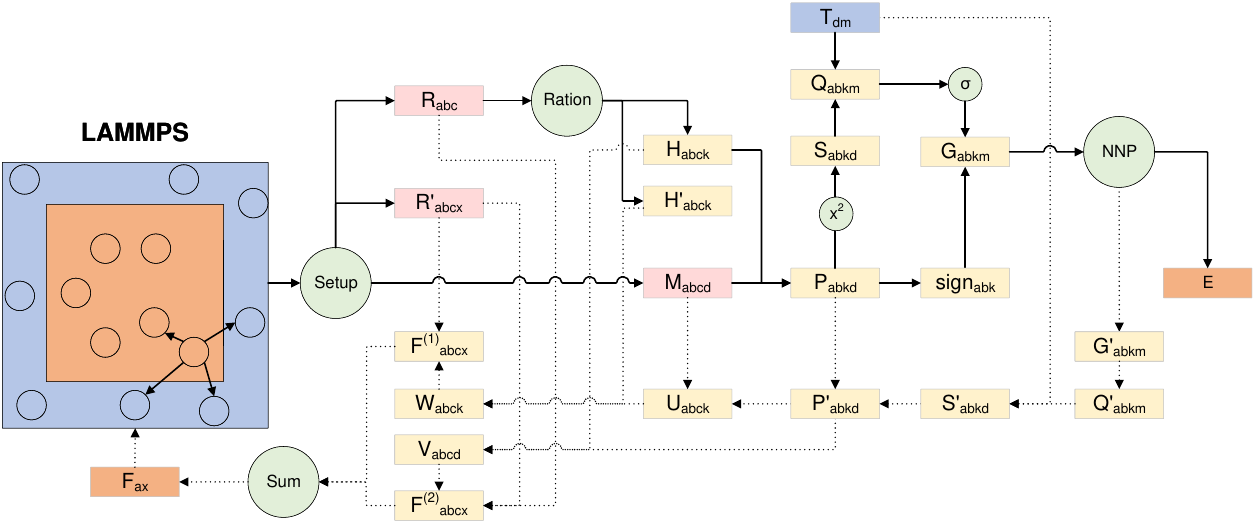}
  \caption{\label{fig:naive} The tensor diagram of TensorMD for the general implementation. Pink tensors are constructed using the setup kernel. Yellow tensors denote intermediate tensors. Orange tensors represent outputs and blue are constant tensors. Green denote kernel functions.}
  \end{center}
\end{figure*}

The associated subscript notations are summarized in Table \ref{table:tensor_notations}.

\begin{table}
  \caption{Tensor subscript notations and definitions. Here \textbf{fixed} indicates whether or not the value will change during simulation. \label{table:tensor_notations}}
  \begin{tabular}{cccc}
    \hline
    Label & Definition & Values & Fixed \\ 
    \hline
    a & The atom index           & 1-100000         & no  \\
    b & The atom pair type index & 1-5              & yes \\
    c & The neighbor index       & 1-1000           & no  \\
    d & The angular moment index & 1,4,10,20 & yes \\
    m & The maximum angular moment & 0-3            & yes \\
    k & The radial interaction kernel index  & 8-128    & yes \\
    x & The Cartesian direction     & 3 & yes \\
    \hline
  \end{tabular}
\end{table}

%

\subsection{General Optimizations}

The comprehensive tensor formalism offers several advantages:

\begin{itemize}
    \item[1.] \textbf{FLOPs and memory operations analysis}: The formalism allows for easy analysis of computational costs, including the number of floating-point operations (FLOPs) and memory operations involved in the computations. This enables further optimizations based on hardware characteristics, leading to potentially more efficient implementations.
    \item[2.] \textbf{Simplified implementation}: The operations involved in the tensor formalism, such as matrix multiplication and batch small matrix multiplication, are relatively straightforward to implement. Utilizing libraries like BLAS (Basic Linear Algebra Subprograms) or MKL (Intel Math Kernel Library) can further enhance the efficiency of the program. Additionally, the well-defined calculations make it easy to incorporate parallel computing techniques such as OpenMP, further improving performance.
    \item[3.] \textbf{Independence from deep learning backends}: The formalism eliminates the need for deep learning backends such as TensorFlow, making it easier to integrate with other programs, such as first-principles codes or other computational chemistry simulations. This allows for greater flexibility and interoperability in the development of atomistic simulation codes.
\end{itemize}

The general version of TensorMD currently relies only on BLAS, with MKL being preferred due to its efficient \textbf{batch\_gemm} APIs (BLAS-like extensions). This simplicity in implementation and independence from deep learning backends make TensorMD a versatile tool for incorporating machine learning techniques into atomistic simulations.

Figure \ref{fig:naive} illustrates the tensor diagram of the entire calculation, showcasing the flow of computations. In addition to the comprehensive tensor formalism, two additional optimizations are included to further enhance the efficiency of the approach.

\subsubsection{The $H_{abck}$ tensor and Batch Rational 1D Interpolation}
The radial interaction tensor $H_{abck}$ can be viewed as applying $k$ independent $\mathbb{R}^1$ to $\mathbb{R}^1$ mappings on $R_{abc}$. Historically, parameterized descriptor functions, such as the Gaussian symmetry function\cite{BPNN} or the exponential function\cite{tensorkmc}, were used to represent these mappings. However, the choice of parameters in these descriptor functions often relied on empirical knowledge, limiting the model extrapolation capabilities. In recent years, deep learning methods have been used to directly model these interactions with neural networks, also known as the embedding net\cite{DPMD_PRL}. The large number of fitting parameters are learned dynamically, leading to better performance. 

TensorMD provides the flexibility of using both parameterized descriptor functions and neural networks for the mapping of $R_{abc}$ to $H_{abck}$. To accelerate the calculation, we adopt an interpolation scheme similar to the tabulation of DP\cite{DP_Ppopp}. Specifically, we use the rational interpolation algorithm proposed by Kerley\cite{Ration}, which is simple and effective. To further speed up the computation, we have developed a batch implementation by pre-computing $s$, $c_1$, and $c_2$. The rational interpolation eliminates the need for the cutoff function $f_c(r_{ij})$ and the mask tensor $\mathrm{mask}_{abc}$, as the smooth effect provided by the cutoff function can be merged into the interpolation. Furthermore, the end points of the interpolation can be forced to zero at $r=0$ and $r\geq r_{cutoff}$, eliminating the need for $\mathrm{mask}_{abc}$. 

We also observed that our batch rational algorithm outperforms direct calculation even for parameterized descriptor functions implemented with highly optimized vectorized libraries, such as the Intel Vector Math Library. The calculation of rational function interpolation involves simple multiplication and division operations, resulting in low computational cost without expensive calculations such as exponential or trigonometric functions. Both $H_{abck}$ and $H_{abck}^{\prime}$ can be represented as matrices with $a \times b \times c$ rows and $k$ columns. Our batch rational interpolation ensures continuous memory access, whereas if parameterized descriptor functions were used and vectorization was desired, loops would have to be carried out over different descriptor functions, resulting in non-continuous memory access.

\subsubsection{The $F^{(2)}_{abcx}$ tensor}
In the original formalism, $F^{(2)}{abcx}$ is calculated by contracting a fourth-order tensor $V{abcd}$ with a fifth-order tensor $M^{\prime}{abcxd}$. This involves small matrix multiplications that are independent of $a$, $b$, and $c$, but $M^{\prime}{abcxd}$ requires a large amount of memory space and is only used once. To overcome this problem, we can implement a more memory-efficient approach: $M^{\prime}_{abcxd}$ is calculated on-the-fly during the contraction instead of pre-calculating and storing it in memory. This reduces the memory footprint (the \textbf{Setup} kernel is significantly accelerated) and eliminates the need for storing and managing large intermediate tensors. 

%

\subsection{Implementation: the new-generation Sunway}

The new Sunway supercomputer is equipped with the brand new SW39000 many-core processors, which are excellent for calculations but have limited memory bandwidth as they only use DDR4 memory (12 channels per CPU). The hardware arithmetic intensity (double precision) of the SW39000 is approximately 43, which is significantly higher than that of Fugaku, a supercomputer based on HBM2 memory, which has an arithmetic intensity of approximately 3. 

\begin{figure*}
  \begin{center}
  \centering
  \includegraphics[width=0.9\textwidth]{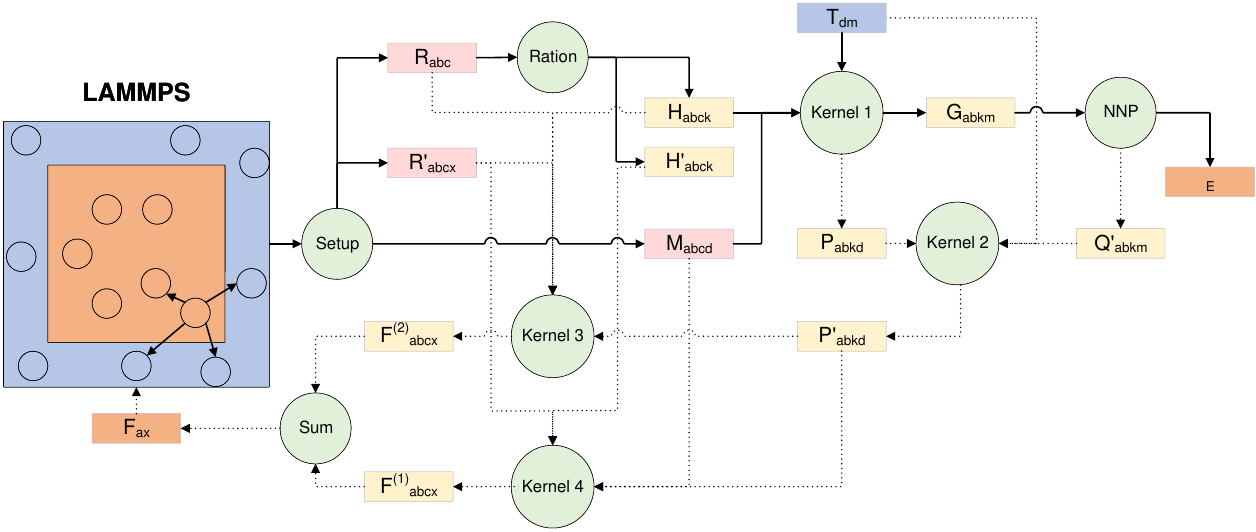}
  \caption{\label{fig:opt_graph_sunway} The tensor diagram of TensorMD potential for Sunway. Pink tensors are constructed using the setup kernel. Yellow tensors denote intermediate tensors. Orange tensors represent outputs and blue are constant tensors. Green denote kernel functions.}
  \end{center}
\end{figure*}

The tensor formalism allows us to easily design kernel fusion strategy (or multiple tensor contractions in a single kernel) to reduce DMA and increase arithmetic intensity. The high speed, fully controllable LDM of SW39000 is the key for designing kernel fusion strategies. However the LDM space is very limited for each CPE. The fusion strategy must be carefully design for maximum performance. Figure \ref{fig:opt_graph_sunway} demonstrates the optimized tensor diagram for SW39000. Total 5 fused kernels were developed:

\begin{itemize}
    \item[1.] The kernel for computing $G_{abkm}$ (Equation \ref{eqn:G_radial_tensor_form}).
    \item[2.] The kernel for computing $P^{\prime}_{abkd}$ (Equation \ref{eqn:dP_abkd}).
    \item[3.] The kernel for computing $F^{(2)}_{abcx}$ with $M^{\prime}_{abcxd}$ calculated on-the-fly.
    \item[4.] The kernel for computing $F^{(1)}_{abcx}$ (Equations \ref{eqn:U_abck} and \ref{eqn:F1_abcx}).
    \item[5.] The kernel for multi-layer dense neural network calculation, both the forward propagation and the backward propagation are implemented.
\end{itemize}

\subsubsection{Fusion Kernels 1-4}

These kernels are similar in that they are all extended batch matrix multiplication kernels. To optimize their performance, we developed an optimization algorithm for many-core processors that utilizes data parallelism. By fusing operators such as batch matrix multiplication (A x B transpose), matrix squaring, dot product, and matrix square root, we reduced redundant DMA operations, and increased data reuse. To address LDM space constraints, we implemented different partitioning strategies for matrix data size in batch matrix multiplication. For small matrix sizes, each sub-core completes a complete fused operator calculation. For larger matrix sizes, we use partially (4) sub-cores to complete the calculation of a complete fused operator calculation, and the data is stored using distributed parallelism. We employ RMA communication to share data and improve memory access bandwidth. For even larger matrix sizes, we use 64 sub-cores to complete a complete fused operator calculation, and adopt core group distributed storage to share data using RMA communication. We fully utilize the LDM local storage to maximize memory access bandwidth.

Moreover, we optimized the core calculation of matrix multiplication with assembly instructions, data padding pre-processing, SIMD, and other techniques, leveraging the computing performance of SW39000. For task parallelism, we implemented double buffering optimization to further improve memory access efficiency. This involves simultaneously calculating the current loop and preparing the necessary data for the next loop with DMA.

\subsubsection{The NNP Kernel}

To enable the computation of multi-layer dense neural networks in a single kernel for the Sunway processor, we have implemented a big-fusion strategy similar to TensorKMC approach. However, TensorMD implementation required significant upgrades to handle the complexity of our situation. Unlike TensorKMC, which uses single-precision floats and only calculates energies during forward propagation, TensorMD uses double-precision floats and requires the derivatives of energy with respect to input for backward propagation. This means that the derivatives of the activations must also be stored in LDM, which is limited for TensorMD. To address this, we adopted a one-to-all RMA broadcast scheme, distributed stored both the weight matrices and their transposes in CPEs evenly, and used the transposed matrices for backward propagation. Additionally, we fused the calculation of $\sigma^{\prime}_{abkm}$ into this NNP kernel.

\subsubsection{The ScatterSum kernel}

This kernel is simple and the plain implementation performs well on Intel CPUs, but using only MPEs on the new generation Sunway will result in extremely poor performance, around 10 times slower. To address this issue, we adopted a sum-reduce strategy. Atomic forces are first summed independently with 8 CPEs (8 threads) and then reduced with all 64 CPEs.

\subsection{Portability}
\label{sec:portability}




The tensor formalism provides a unified framework for representing various interatomic potentials. TensorMD can be viewed as a special case of BPNN and it also originates from MEAM. This suggests that the complicated MEAM potential can also be expressed using tensor diagrams, which will be presented in another paper. The DP model can be represented with our tensor framework as well. The tensor formalism also eliminates the need for deep learning backends such as TensorFlow or PyTorch, making it more accessible to researchers in different fields.

Moreover, algorithmic innovations of TensorMD, including the portable batch rational interpolation and batch matrix multiplication kernels, have potential applications in various scientific domains beyond atomistic materials science (for example the rational interpolation is widely used in hydrodynamics simulations as its the fundamental interpolation algorithm for OpenSesame EOS). These innovations are also easier to integrate with other scientific codes due to their portability.

The parallelization strategies and optimization techniques used in TensorMD are also transferable to other scientific codes, allowing for improved performance on modern high-performance computing architectures. Overall, the innovations presented in TensorMD have the potential to accelerate scientific research in multiple fields beyond materials science, making the tensor formalism a valuable tool for a wide range of scientific applications.

%

\section{Evaluation}

%

\subsection{The Potential}

In this work, we trained two TensorMD potentials for W using the open-source program TensorAlloy\cite{TensorAlloy_1}. The small version of the potential has 32 radial interaction filters and 3 hidden layers and was designed for studying dynamic behaviors. The large version has 128 filters and 4 hidden layers and was optimized for free energy studies. The dataset used for training consisted of 18,000 structures covering a wide range of thermodynamic conditions (0-2000 GPa, 0-14,000 K) and was obtained from first-principles calculations using the PBE\cite{PBE} functional and VASP\cite{VASP}. The potentials specifications and relative errors (current version) are presented in Table \ref{table:pot_mae}, and the SquarePlus activation\cite{barron2021squareplus} function was used to avoid logarithm and exponential calculations. Both potentials were found to accurately predict elastic constants with absolute errors within 2 GPa compared to DFT. A comprehensive evaluation of the two potentials and their applications in high-pressure phase transformation research will be presented in a forthcoming paper.

\begin{table}
  \caption{Performances of the potentials \label{table:pot_mae}}
  \begin{tabular}{ccc}
    \hline
     & $k=32$ & $k=128$ \\ 
    \hline
    NNP & 128,128,128,1 & 128,128,128,128,1  \\
    Activation & SquarePlus & SquarePlus \\
    Energy MAE (meV/atom) & 5.9       & 4.1  \\
    Forces MAE (eV/\AA) & 0.09       & 0.08  \\
    Stress MAE (GPa) & 0.55       & 0.51  \\
    \hline
  \end{tabular}
\end{table}

%
\subsection{Performance Results for the General Implementation}

In this section, we will discuss the performance results of the general version of our program on a node equipped with two Intel Xeon Gold 6336Y processors, each with 28 cores running at 2.4 GHz. The total memory available on this node is 384 GB (12 channels), and the theoretical maximum memory bandwidth is 307 GB/s. To achieve maximum performance, we used the Intel Math Kernel Library (version 2022.2). 

\subsubsection{The Batch Rational Interpolation}

We performed two types of tests to assess the accuracy of our interpolation algorithm with respect to $\Delta r$: a static test and a dynamic test. In the static test, we computed atomic forces for the same configuration (16000 atoms) using different $\Delta r$ values and calculated the root mean squared errors (RMSE). The dynamic test involved running molecular dynamics (MD) simulations using the same initial configuration (54000 atoms) and comparing the RMSE of thermodynamic metrics, such as total energy, temperature, and pressure. The reference results are obtained by directly calculating the neural network $H_{abck}$.

\begin{figure*}
  \begin{center}
  \centering
  \includegraphics[width=1\textwidth]{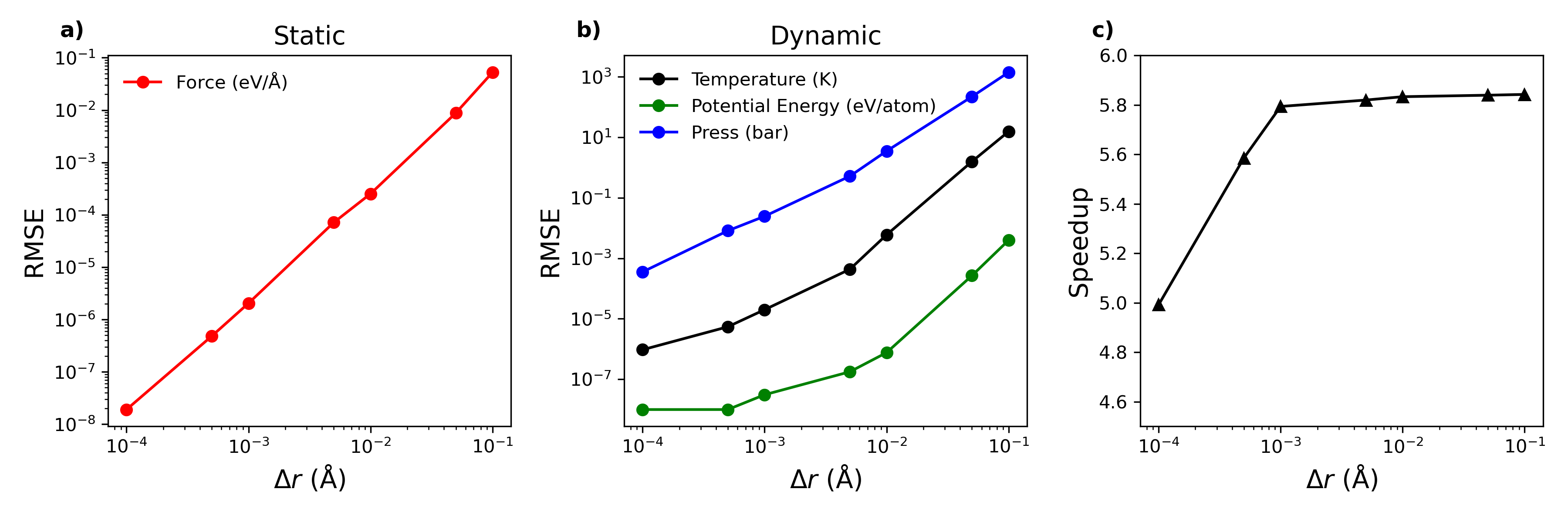}
  \caption{\label{fig:ration_accuracy_speed} The static \textbf{a)} and dynamic \textbf{b)} RMSE errors and \textbf{c)} speed up with respect to $\Delta r$ for the batch rational interpolation. }
  \end{center}
\end{figure*}

Figure \ref{fig:ration_accuracy_speed} displays the results for RMSE and speedup with respect to $\Delta r$. The speedup results were obtained by simulating 54,000 atoms with 1,000 steps and 48 MPI processes under the NVT ensemble and ambient condition. Our experiments suggest that the optimal range for $\Delta r$ is between 0.01 \AA{} and 0.001 \AA{}, which is consistent with the findings of DP\cite{DP_Ppopp}. In general, a value of $\Delta r = 0.01$ \AA{} may be preferred for the general implementation, while $\Delta r = 0.01$ \AA{} is still acceptable for heterogeneous systems.

\begin{figure}
  \begin{center}
  \centering
  \includegraphics[width=0.4\textwidth]{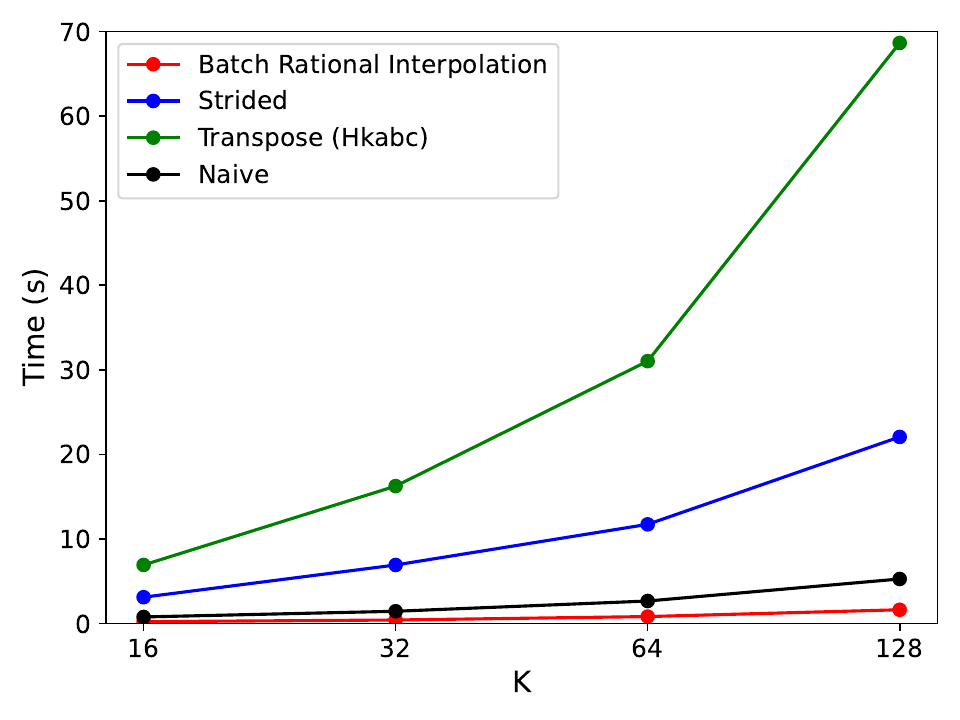}
  \caption{\label{fig:ration_test} Performance results of different algorithms for computing parameterized functions based $H_{abck}$}
  \end{center}
\end{figure}

Figure \ref{fig:ration_test} compares the performance of batch rational implementation to other approaches for computing parameterized descriptor functions, using the exponential descriptor function as implemented by Oganov\cite{Oganov_U} and TensorKMC\cite{tensorkmc}. The results were obtained using standalone unittest, 48 MPI processes and $a = 2000, b = 1, c = 200$ for each process. The "naive" approach refers to a basic double for loop implementation, while "strided" is implemented using strided Intel Vectorized Math Library (VML) APIs. The "Transpose($H_{kabc}$)" approach is a two-step strategy that involves computing $H_{kabc}$ with continuous VML APIs and then performing an in-place transpose using MKL. The results suggest that memory continuity is quite important. The batch rational implementation is the fastest, more than 3 times faster than the naive implementation. Hence, we recommend that other codes use this interpolation algorithm for similar calculations.

\subsubsection{The $F^{(2)}_{abcx}$ with $M^{\prime}_{abcxd}$ calculated on-the-fly}

\begin{table}
  \caption{Overall and kernel time (seconds) before and after applying the fused $F^{(2)}_{abcx}$ strategy \label{table:Fused_F2}}
  \begin{tabular}{ccc}
    \hline
            & Original & Optimized \\ 
    \hline
    Setup   & 29.08  & 8.42  \\
    F2      & 8.71   & 5.96  \\
    Overall & 130.86   & 107.51  \\
    \hline
  \end{tabular}
\end{table}

Table \ref{table:Fused_F2} provides a summary of the time difference before and after applying the fused $F^{(2)}_{abcx}$ strategy. The time results were obtained by simulating 54,000 atoms with 1,000 steps and 48 MPI processes under the NVT ensemble and ambient condition. The application of this simple strategy had a significant impact on the performance. The kernel setup time was reduced by \~75\%, and surprisingly, the F2 kernel itself was faster after the optimization.

\subsubsection{Overall}

Until now, there have been several MLIP benchmark tests on CPUs. However, many of these tests used only one core, which may not be appropriate since MLIPs are typically memory-bound. Table 1 summarizes the benchmark tests of the TensorMD general implementation and some other references. Surprisingly, the fully optimized TensorMD using the small potential is quite close in performance to semi-empirical potentials like MEAM, demonstrating the capability of TensorMD in achieving high efficiency in simulating large-scale systems.

However, it is worth noting that benchmark tests on different systems and with different potentials may yield different results. Therefore, it is important to carefully evaluate the performance of MLIPs on a specific system and under specific conditions before making any conclusions. Additionally, future developments in hardware and software may also affect the performance of MLIPs, so regular benchmarking and optimization are necessary to ensure their efficiency and accuracy.

\begin{table}
  \caption{Benchmark results on our Intel X86 computing node, ToS means time-to-solution ( $\mathbf{\mu s}\cdot$core/step/atom).}
  \begin{tabular}{cccccc}
    \hline
    Potential & System & \# Atoms & \# MPI & ToS \\ 
    \hline
    DP-SE2\cite{DP_W}   & W      & 128    & 1   & 657  \\
    DP-HYB\cite{DP_W}   & W      & 128    & 1   & 1475 \\
    GAP-2\cite{DP_W}    & W      & 128    & 1   & 4585 \\
    DP-SE2\cite{DP_W}   & Cu     & 6912   & 6   & 123  \\
    MEAM\cite{W_MEAM}   & W      & 128    & 1   & 16   \\
    MEAM\cite{W_MEAM}   & W      & 6912   & 6   & 14   \\
    MEAM\cite{W_MEAM}   & W      & 55296  & 48  & 15   \\
    This ($k=32)$       & W      & 128    & 1   & 34   \\
    This ($k=32)$       & W      & 6912   & 6   & 42   \\
    This ($k=32)$       & W      & 55296  & 48  & 96   \\
    This ($k=128)$      & W      & 1      & 1   & 101   \\
    This ($k=128)$      & W      & 6912   & 6   & 98   \\
    This ($k=128)$      & W      & 55296  & 48  & 282   \\
    \hline
  \end{tabular}
\end{table}

%
\subsection{Performance Results for the Sunway Implementation}

\subsubsection{The Arithmetic Intensity Analysis}

The tensor formalism simplifies the estimation of arithmetic intensity. In this section, we analyze the theoretical arithmetic intensity of TensorMD and its individual kernels at varying input sizes of $c$ and $k$. $c$ varies significantly with different densities (pressures). The computational intensity is defined as the ratio of floating-point operations to memory reads/writes (flops:byte ratio), and only DMA memory operations are considered in this analysis. If the computational intensity exceeds the Maximum Arithmetic Intensity of the hardware, the program is considered compute-bound and can fully utilize the computing power of the hardware. The neural network size is fixed at [128, 128, 128, 1].

\begin{figure}
  \begin{center}
  \centering
  \includegraphics[width=0.45\textwidth]{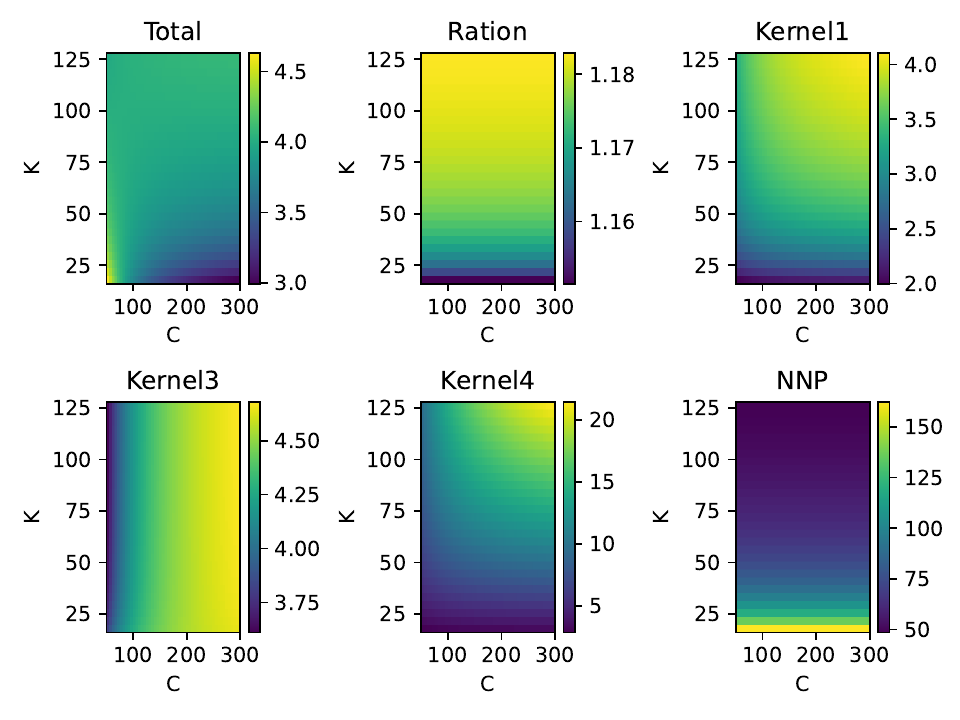}
  \caption{\label{fig:arithmetic_intensity} The arithmetic intensity with respect to $c$ and $k$ of the Sunway implementation. }
  \end{center}
\end{figure}

As shown in Figure \ref{fig:arithmetic_intensity}, the overall computational intensity of TensorMD reaches its maximum value of 4.631 at c=50 and k=16. Even under extreme conditions, the computational intensity of TensorMD is much lower than the maximum computational intensity of the SW39000 (~44), indicating that the program is significantly memory-bound and may not fully utilize the computing power of the new generation of the Sunway supercomputer.

The individual kernels in TensorMD were also analyzed, including Ration, Kernel1, Kernel3, Kernel4, and NNP. As shown in the figure, the arithmetic intensity of Ration reaches its maximum value of 1.183 at k=128. The arithmetic intensity of Kernel1 reaches its maximum value of 4.105 at c=300 and k=128. The arithmetic intensity of Kernel3 reaches its maximum value of 4.68 at c=300 and k=128. The arithmetic intensity of Kernel4 reaches its maximum value of 21.47 at c=300 and k=128. The neural network kernel reaches its maximum arithmetic intensity value of 162.5 at k=16.

Most of the kernels in TensorMD have low computational intensity and are memory-bound, but the NNP kernel has high computational intensity in theory and brings a large optimization potential.

\subsubsection{Step-by-step Results}

This section presents the step-by-step optimization results analysis. The general implementation built with the Sunway xMath library is used as the baseline. All optimized kernels were enabled in TensorMD step by step while recording the loop time and speedup of 100 steps simulation. The test was conducted under two different potentials($k=32$ and $k=128$), lattice constant 3.1887 \AA \  and temperature 300 K. The simulation box has 85750 atoms for $k=32$ and 23040 atoms for $k=128$. One SW39000 processor (6 CGs) is used to measure the performance. 

\begin{figure*}
  \begin{center}
  \centering
  \includegraphics[width=0.9\textwidth]{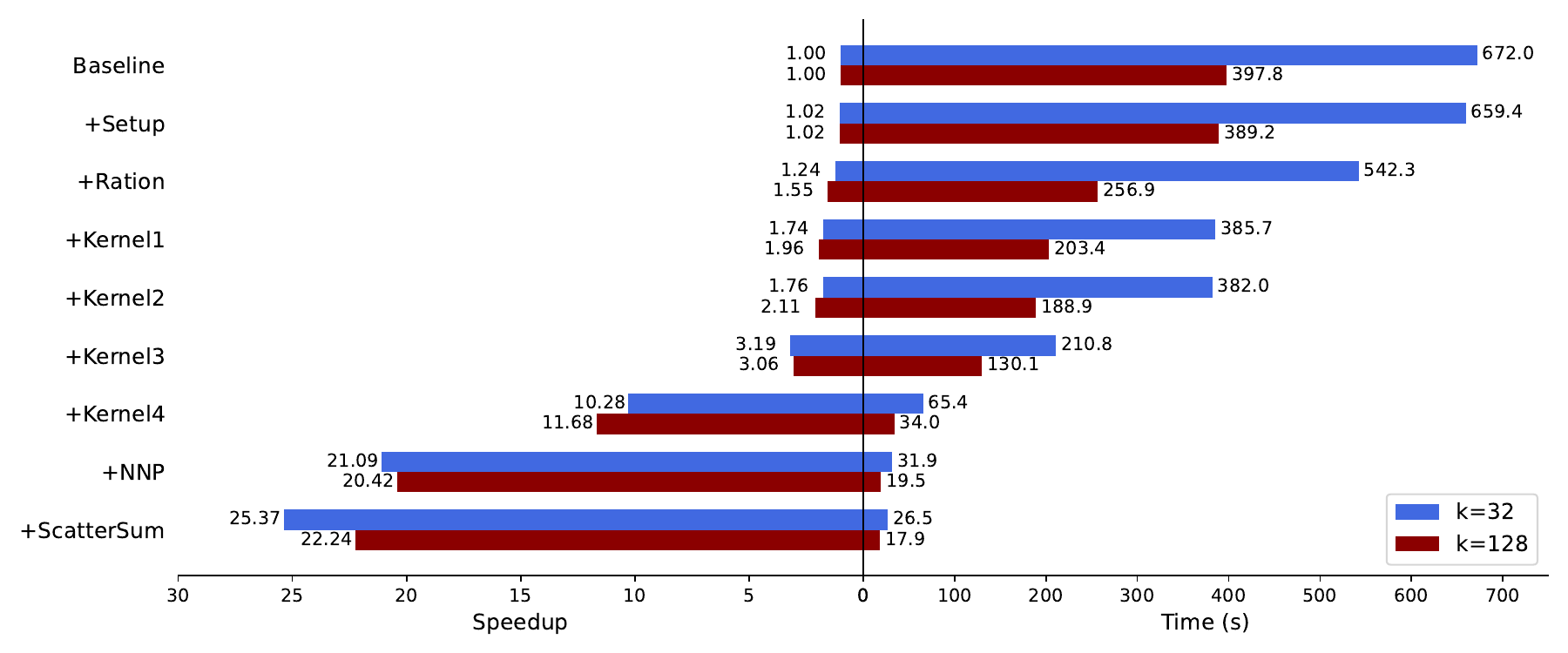}
  \caption{\label{fig:step_by_step} Step-by-step results on Sunway: \textbf{(left)} the speed up factors and \textbf{(right)} the absolute changes in seconds. + denotes the point at which the optimized kernel was first used.}
  \end{center}
\end{figure*}

As shown in Figure \ref{fig:step_by_step}, all optimized kernels bring performance improvement, among which Ration, Kernel1, Kernel3, kernel4 and NNP have the most significant effect. The performance improvement of Setup and ScatterSum comes from slave core athread optimization. Ration benefits from slave core athread, DMA and SIMD optimization. And the improvement of Kernel 1-4 and NNP is mainly due to DMA optimization and elimination of intermediate variables that reduce the number of memory access. The acceleration ratio of the NNP kernel is not that significant compared with the big-fusion operator for TensorKMC due to many reasons: single-to-double precision floats (equivalent memory bandwidth halved ), much more RMA broadcasts and much more DMA write (from $a$ singles to $(a \times b \times k \times (m + 1) + a)$ doubles).

Finally, the combination of optimized kernels brings TensorMD the maximum speedup of 25.37.

Table \ref{table:sunway_intel_comparison} provides a comparison of the performance between the Sunway implementation and the general implementation. The results suggest that one SW39000 processor is roughly equivalent to 1.6 Intel Xeon Gold 6336Y processors in terms of performance.  

\begin{table}
  \caption{Performance comparison (seconds) between the Sunway implementation and the general implementation for TensorMD \label{table:sunway_intel_comparison}}
  \begin{tabular}{ccccc}
    \hline
    Platform & \# MPI & \# OpenMP & $k=32$ & $k=128$  \\ 
    \hline
    Sunway   & 6      &           & 26.5   & 17.8     \\
    \hline
    Intel    & 6      & 1         & 61.6   & 47.0     \\
             & 6      & 8         & 22.0   & 15.9     \\
             & 48     & 1         & 18.0   & 13.3     \\
    \hline
  \end{tabular}
\end{table}

%
\subsection{Scalability Results}

This section presents the scalability testing of the optimized TensorMD on the new Sunway supercomputer for large-scale simulations of bulk W systems ranging from 20 million atoms to a maximum of 52 billion atoms, representing the largest MLIP based MD simulations to date. Each simulation runs for 50 steps, during which thermal dynamics properties, including total energy, temperature, kinetic energy, pressure, volume, and the diagonal components of the pressure tensor (pxx, pyy, pzz), are collected and recorded every 10 steps. The neighbour list is updated using the binning method with a 1 Å buffer region every 5 steps. The NVT ensemble is used with a temperature of 300 K and a time step of 1 fs. These are commonly used settings for MD simulations, particularly with respect to the neighbor list update frequency and thermo update frequency, making our scalability results applicable to real-world situations. Loop time is used to measure performances.

\subsubsection{Strong Scaling}

\begin{figure}
  \begin{center}
  \centering
  \includegraphics[width=0.4\textwidth]{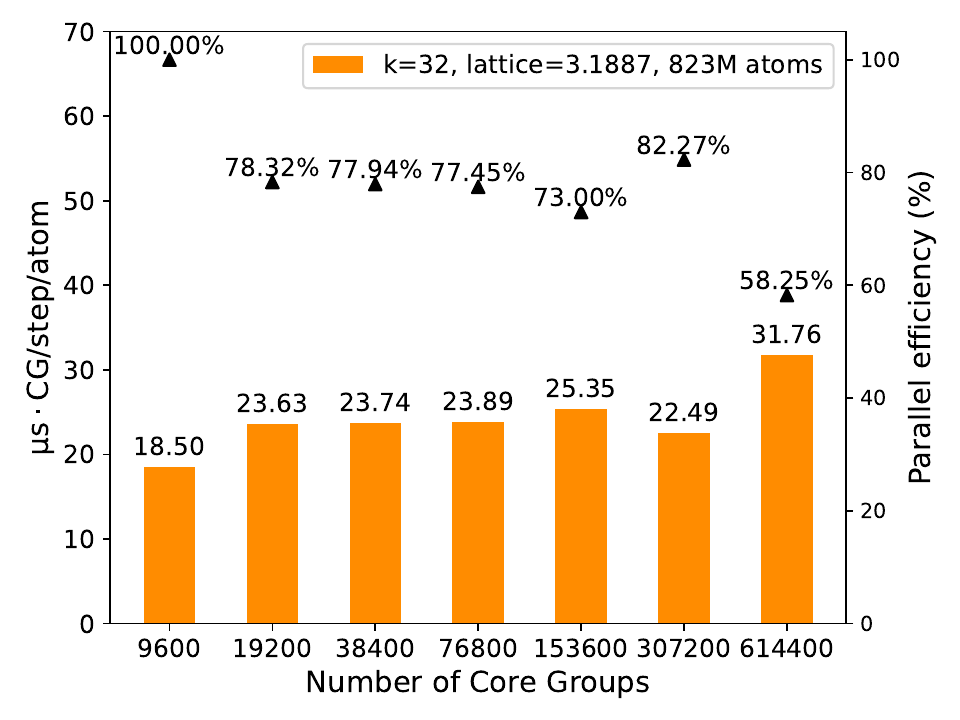}
  \caption{\label{fig:strong_scaling} The strong scaling results on Sunway. }
  \end{center}
\end{figure}

The strong scaling test is performed with the $k=32$ potential at ambient condition (lattice constant 3.1887 \AA{} and temperature 300 K), starting from 9,600 core groups and scaling up to a maximum of 614,400 core groups, with total 823.3 million atoms. The number of atoms per CG ranges from 85750 to 1340. As shown in Figure \ref{fig:strong_scaling}, TensorMD demonstrated excellent strong scalability, maintaining 58.25\% parallel efficiency even with 614,400 core groups. This result also indicates that the node communication performance of the Sunway supercomputer is excellent and suitable for large-scale molecular dynamics simulations.

\begin{figure}
  \begin{center}
  \centering
  \includegraphics[width=0.4\textwidth]{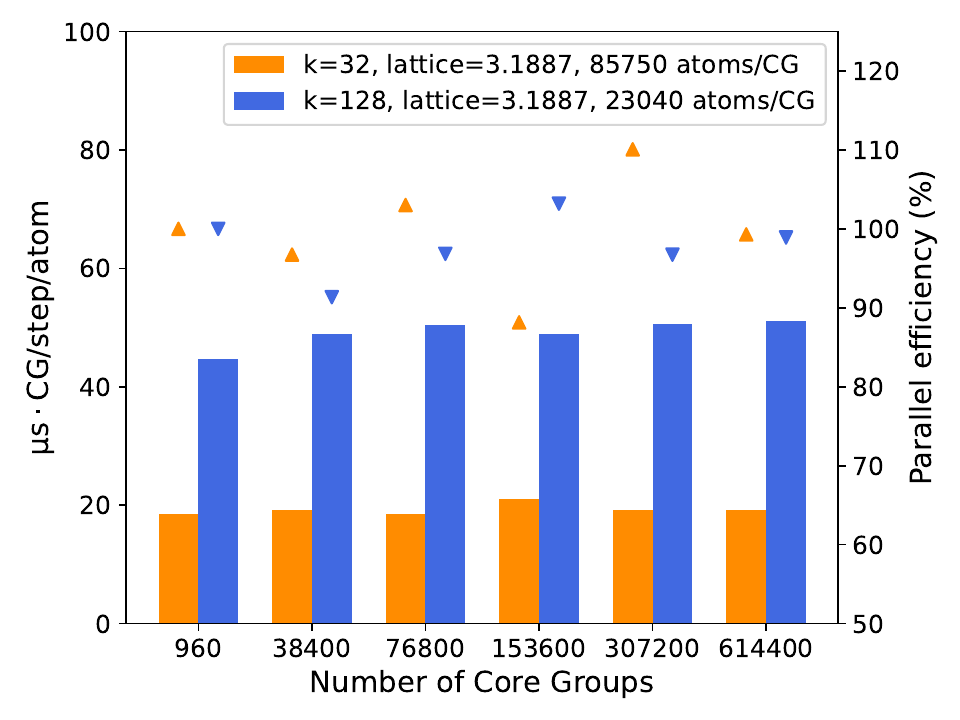}
  \caption{\label{fig:weak_scaling_1} The weak scaling results with lattice constants 3.1887 \AA{} on Sunway. }
  \end{center}
\end{figure}

\begin{figure}
  \begin{center}
  \centering
  \includegraphics[width=0.4\textwidth]{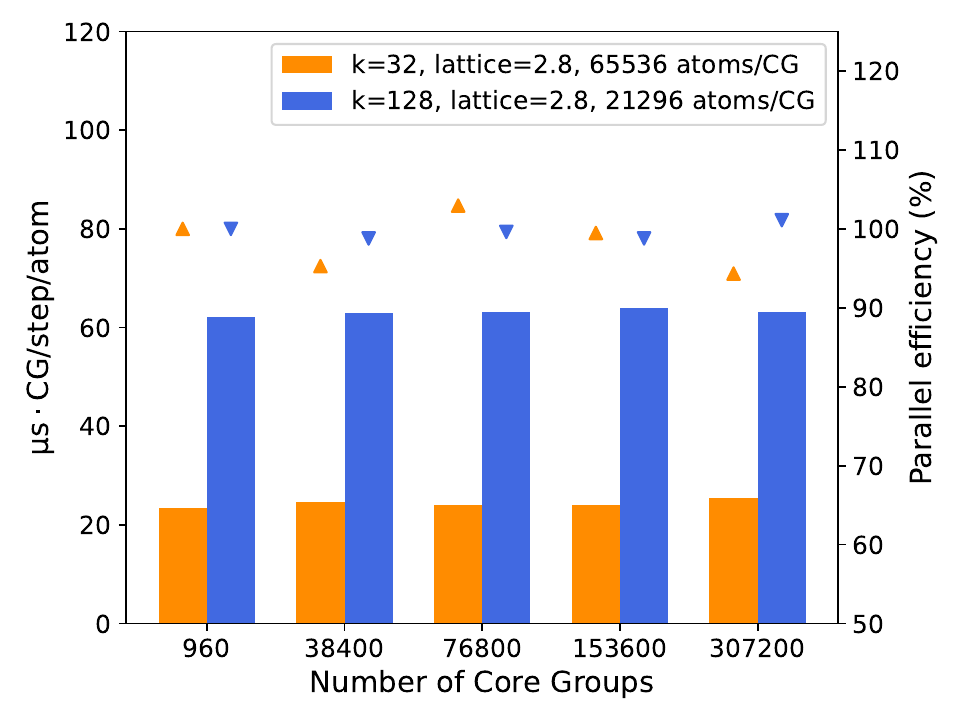}
  \caption{\label{fig:weak_scaling_2} The weak scaling results with lattice constants 2.8 \AA{} on Sunway. }
  \end{center}
\end{figure}

\subsubsection{Weak Scaling}

We conducted four weak scaling tests using two different potentials ($k=32$ and $k=128$) and two lattice constants (3.1887 \AA{} and 2.8 \AA). 3.1887 \AA{} corresponds to 0 GPa while 2.8 \AA{} is approximately 270 GPa. Lowering the lattice constant increases the material density, which leads to a significant increase in computational overload as the number of neighbors grows cubically due to the fixed cutoff radius in MLIP. The tests started with a baseline of 960 cores and scaled up to a maximum of 307,200 cores, with the number of atoms per core group set according to the maximum simulation size for that environment in the table, up to a maximum of 26.3424 billion atoms. As shown in the results, TensorMD demonstrated excellent weak scalability as the number of core groups increased. Furthermore, it can be observed that decreasing the lattice scale and increasing the number of descriptor functions both lead to an increase in simulation time per atom, with the latter having a more significant impact. 

We also conducted two additional full-machine tests to compare with previous state-of-the-art results at ambient conditions with a lattice constant of 3.1887 \AA. Using 614,400 core groups, TensorMD was able to simulate up to 52.6 billion atoms, which is currently the largest MLIP-based MD simulation ever performed, with a time-to-solution of $3.1 \times 10^{-11}$ s/step/atom, setting a new record \ref{table:ToS}. 

\begin{table}
  \caption{Time-to-solution (ToS, s/step/atom) results (double precision) of DP, SNAP and this work. \label{table:ToS}}
  \begin{tabular}{ccccc}
    \hline
    MLIP & System & \# Atoms & Platform & ToS \\ 
    \hline
    DP\cite{DP_GB}          & Cu & 127M & Summit & $8.1 \times 10^{-8}$  \\
    SNAP\cite{snap_gb_sc21} & C  & 20B  & Summit & $3.5 \times 10^{-11}$ \\
    DP\cite{DP_Ppopp}       & Cu & 3.4B & Summit & $1.1 \times 10^{-10}$ \\
    DP\cite{DP_Ppopp}       & Cu & 17B  & Fugaku & $4.1 \times 10^{-11}$ \\
    This ($k=128$)          & W  & 14B  & Sunway & $8.3 \times 10^{-11}$ \\
    This ($k=32$)           & W  & 52B  & Sunway & $3.1 \times 10^{-11}$ \\
    \hline
  \end{tabular}
\end{table}

The excellent weak scalability suggests that TensorMD is capable of performing even larger-scale material simulations in the future as hardware improves, particularly in terms of memory.

%
\section{Conclusions}

Molecular dynamics simulations have become increasingly important in materials science due to their effectiveness in predicting the behavior of complex systems. One area of growing interest is machine-learning interatomic potentials (MLIPs), which use machine learning techniques to model interatomic interactions. In this work, we introduced a new MLIP model named TensorMD, which is based on a physical background and utilizes a tensor formalism. The tensor formalism allows for more efficient computation and greater flexibility to work with other scientific codes. We proposed several portable optimization strategies and developed a highly optimized version for the new Sunway supercomputer. TensorMD can simulate up to 52 billion atoms with a time-to-solution of 31 ps/step/atom, setting new records for HPC + AI + MD.

The current Sunway implementation can already achieve 2-4 times greater cost-efficiency for large-scale molecular dynamics simulations compared to Intel/AMD supercomputers, estimated with public core$\cdot$hour prices and our high-pressure phase transformation studies of W.
Such economic efficiency can allow researchers to simulate larger and longer, which may lead to new scientific discoveries in materials science.

%

\bibliographystyle{unsrt}
\bibliography{./main}

\end{document}